\documentclass{vienna-conf2019}
\usepackage{graphicx}
\usepackage{hyperref}
\usepackage[]{natbib}
\usepackage{epstopdf}
\usepackage{color}
\usepackage[normalem]{ulem} 

\def\BibTeX{{\rm B\kern-.05em{\sc i\kern-.025em b}\kern-.08em
    T\kern-.1667em\lower.7ex\hbox{E}\kern-.125emX}}
\bibpunct{(}{)}{;}{a}{}{,}  

\begin{document}

\TitreGlobal{Stars and their variability observed from space}


\title{Gaia's revolution in stellar variability}

\runningtitle{Gaia's revolution in stellar variability}

\author{L. Eyer}\address{Department of Astronomy, University of Geneva, Chemin des Maillettes 51, 1290 Versoix, Switzerland}

\author{L. Rimoldini}\address{Department of Astronomy, University of Geneva, Chemin d'Ecogia, 1290 Versoix, Switzerland}



\author{L. Rohrbasser$^{2}$}
\author{B. Holl$^{1}$}
\author{M. Audard$^{1,2}$}
\author{D.W. Evans}\address{Institute of Astronomy, University of Cambridge, Madingley Road,
10 Cambridge CB3 0HA, UK}

\author{P. Garcia-Lario}\address{European Space Astronomy Centre (ESA/ESAC), Villanueva de la Canada, 28692 Madrid, Spain}
\author{P. Gavras$^{4}$}
\author{G. Clementini}\address{Osservatorio di Astrofisica e Scienza dello Spazio di Bologna, Via Piero Gobetti 93/3, 40129 Bologna, Italy}
\author{S. T. Hodgkin$^{2}$}
\author{G. Jevardat de Fombelle$^{2, }$}\address{SixSq, Route de Meyrin 267, 1217 Meyrin, Switzerland}
\author{A. C. Lanzafame}\address{Università di Catania, Dipartimento di Fisica e Astronomia, Sezione Astrofisica, Via S. Sofia 78, 95123 Catania, Italy}\secondaddress{Osservatorio Astrofisico di Catania, Via S. Sofia 78, 95123 Catania, Italy}
\author{T. Lebzelter}\address{University of Vienna, Department of Astrophysics, Türkenschanzstraße 17, 1180 Vienna, Austria}
\author{I. Lecoeur-Taibi$^{2}$}
\author{N. Mowlavi$^{1,2}$}
\author{K. Nienartowicz$^{2, 7}$}
\author{V. Ripepi}\address{Osservatorio Astronomico di Capodimonte, Via Moiariello 16, 80131, Napoli, Italy}
\author{L. Wyrzykowski}\address{Astronomical Observatory, University of Warsaw, Al. Ujazdowskie 4, 00-478 117 Warszawa, Poland}

\setcounter{page}{237}


\maketitle


\begin{abstract}
Stellar variability studies are now reaching a completely new level thanks to ESA's Gaia mission, which enables us to locate many variable stars in the Hertzsprung-Russell diagram and determine the various instability strips/bands. Furthermore, this mission also allows us to detect, characterise and classify many millions of new variable stars thanks to its very unique nearly simultaneous multi-epoch survey with different instruments (photometer, spectro-photometer, radial velocity spectrometer).
An overview of what can be found in literature in terms of mostly data products by the Gaia consortium is given. This concerns the various catalogues of variable stars derived from the Gaia time series and also the location and motion of variable stars in the observational Hertzsprung-Russell diagram.
In addition, we provide a list of a few thousands of variable white dwarf candidates derived from the DR2 published data, among them probably many hundreds of new pulsating white dwarfs. On a very different topic, we also show how Gaia allows us to reveal the 3D structures of and around the Milky Way
thanks to the RR~Lyrae stars.
\end{abstract}



\section{Introduction}
The Time-Domain revolution for ``large" sky surveys {\it from space} keeps advancing strongly since the 1990s, with the Hipparcos mission \citep{Hipparcos1997}, 
CoRoT \citep{COROT2003}, and the Kepler/K2 survey \citep{Kepler2010,K22014}.
Today, we are experiencing a major advance
in global space-based surveys thanks to Gaia \citep{Gaia2016} and TESS \citep{TESS2015}.
However, we are still at the stage where each space mission is unique, and we can claim that Gaia benefits from very unique features. One obvious feature is the exceptional astrometric precision, but this is not the only feature. Its (spectro)photometric survey with its G band photometry, integrated BP and RP photometry, and BP and RP low resolution spectra is very singular. Gaia remains unique even without its photometry and astrometry, thanks to its spectroscopic survey that covers more than one hundred million objects \citep{Sartoretti2018}. In addition, since the entire sky is measured with one set of instruments and measurements are nearly simultaneous, the physical interpretation of the data is facilitated. The uniqueness of Gaia is also revealed from its stunning numbers accumulated during 5 years: nearly 1.3 trillions of astrometric CCD measurements, close to 1.5 trillion G-band/BP/RP photometric CCD data and more than 25~billion spectroscopic measurements.

From all these data of different nature, we can perform large-scale statistical descriptions of the stellar populations of variable stars in the Milky Way and its neighbourhood.

\section{A few examples of Gaia results}

\subsection{Science alerts}
While scanning the entire sky, Gaia continuously reports alerts\footnote{\url{http://gsaweb.ast.cam.ac.uk/alerts/home}} on astronomical events whose science could be missed if they were not followed up promptly by the community. As of mid-November 2019, the Gaia Photometric Science Alerts team have released more than 10,000 alerts, at a current rate of more than 10 alerts per day. About 25\% of the alerts are classified, and out of these about two-thirds turn out to be Supernovae (e.g., \citealt{Kangas2017}). Gaia has also discovered almost 300 Cataclysmic Variables, and perhaps most excitingly the eclipsing AM CVn system Gaia14aae (\citealt{Campbell2015}). Thanks to significant community involvement, Gaia has also found in excess of 30 confirmed microlensing events (from $>$100 candidates, most of which await confirmation), located mostly outside of the Galactic bulge (e.g. \citealt{Wyrzykowski2019}).

\subsection{Variable stars in the first Gaia Data Releases}

Variable stars data were published in the first Gaia data release. The release was small since its main goal was to demonstrate Gaia's photometric capabilities, and to expose the general approach/methodology to analyse the photometric time series of Gaia \citep{Eyer2017}. We released 3,194 G-band time series of Cepheids and RR~Lyrae stars from the first 14 month solution and concentrated on stars sampled during the ecliptic pole scanning law \citep{Clementini2016}. There was also a science demonstration article on the Period Luminosity (PL) relations of Cepheids and infrared PL and optical luminosity-metallicity relations of RR Lyrae stars \citep{Clementini2017} calibrated on parallaxes of the Tycho-Gaia Astrometric Solution (TGAS, \citealt{Lindegren2016}).

\subsection{Variable stars in the second Gaia Data Releases }

As part of the second data release,  ``only" half a million stars with Gaia variability information were published \citep{Holl2018}. The resulting catalogues are among the largest catalogues of RR~Lyrae stars \citep{Clementini2019,Rimoldini2019}, $\delta$ Scuti and SX Phoenicis stars \citep{Rimoldini2019}, Long Period variable stars \citep{Mowlavi2018,Rimoldini2019}, rotationally modulated variable stars \citep{Lanzafame2018,Lanzafame2019}. We further published a catalogue of Cepheids in the Galaxy and the Magellanic Clouds \citep{Clementini2019,Rimoldini2019,Ripepi2019} and of short-timescale ($< 1$ day) variables \citep{Roelens2018}. For all these sources, more than 1.6 million photometric time series (G-band, integrated BP and RP bands) are published as well as statistical attributes and specific parameters for specific groups.

Another product is the science demonstration article  \citep{Eyer2019}, where we show where stars of certain variability types are located and how they move in the observational Hertzsprung-Russell diagram. This description is totally unprecedented. This was done for pulsating stars, cataclysmic stars, eruptive, eclipsing binaries/exo-planet transit hosts, and rotational-induced variable stars. 
For cataclysmic variable stars, the samples have been cleaned and sub-classified in \cite{Pala2019}
and \cite{Abril2019}. We also showed how the stars are moving in the Hertzsprung-Russell diagram in several
movies, e.g. on  YouTube\footnote{\url{https://www.youtube.com/watch?v=Pcy4U5uvL8I}}. 

We also derived the fraction of variable star as detected with the Gaia precision of DR2 over the entire Hertzsprung-Russell diagram.
It is quite striking that, in the instability strip at the location of $\delta$ Scuti stars, the fraction of variable stars is not so different than the one of \cite{Murphy2019} based on Kepler data. 

We were also surprised to see the fraction of variables along the white dwarf sequence which prompted us to analyse further this region of the Hertzsprung-Russell diagram using the DR2 published data (see Section~\ref{sec:WDvar}).

\subsection{Images of the week}
There are also examples which are published in the Gaia Image of the Week on the ESA webpage\footnote{\url{https://www.cosmos.esa.int/web/gaia/image-of-the-week}}, which demonstrate the capabilities of Gaia on many different topics, 
including features of time series of photometry/spectroscopy. In that way astronomers can get an early view of what Gaia is capable of. A few of them are mentioned below.
\begin{itemize}
 \item 29/05/2019: the RVS spectra time series of X Per (a member of the Be/X-ray class of binaries) is visualised by an animation that shows clear variability of the emission lines of the Be star, following the accretion of the circumstellar matter around the Be star on the neutron star companion. The animation of spectra time series of other stars are shown, for example one related to the RS CVn system SZ Psc which shows strongly variable chromospheric activity in the Calcium lines.
  \item 24/05/2019: the period-amplitude diagram of rotational modulation variable candidates shows a multi-modal structure that is interpreted as different regimes in surface inhomogeneities as function of age, see \cite{Lanzafame2019}.
  \item 18/12/2018: the outburst of a rare FU Orionis star showed the extreme changes in brightness and spectral type that are typical of this type of young stellar objects.
  \item 15/11/2018: the epoch spectra in RP were used to determine if the atmosphere of an evolved star is carbon or oxygen rich \citep[as also described in][]{2019gaia.confE..62M}.
  \item 24/03/2017: the median colour-magnitude diagram of fundamental mode RR~Lyrae stars was shown to be able to identify the location of these variable stars, from the galactic halo to the direction to the bulge, from the Sagittarius dwarf spheroidal galaxy to its tidal streams, and from the Large to the Small Magellanic Cloud.
 \item 09/10/2015: the radial velocity of a spectroscopic double-lined binary (SB2) was derived for each binary component as a function of time from the epoch spectra of Gaia's Radial Velocity Spectrometer (RVS).
\end{itemize}

\subsection{Other results on variable stars}
The Gaia data has been used for many purposes in relation to variable stars. Here, we highlight a few studies that developed a virtuous circle between Gaia and these topics.  We have a stunning improvement of the RR Lyrae luminosity properties \citep{Muraveva2018} thanks to the high relative precision of Gaia parallaxes for a large number of stars. There is an an interplay with other distance-determination methods based on asteroseimology \citep{Zinn2019}, Cepheids \citep{Riess2018}, and eclipsing binaries \citep{Graczyk2019}, although all of these studies have a larger parallax offset term than the one resulting from quasars \citep{Lindegren2018,Arenou2018}.

\section{New pulsating white dwarf candidates\label{sec:WDvar}}
\begin{figure}[ht!]
 \centering
 \includegraphics[width=0.799\textwidth,clip]{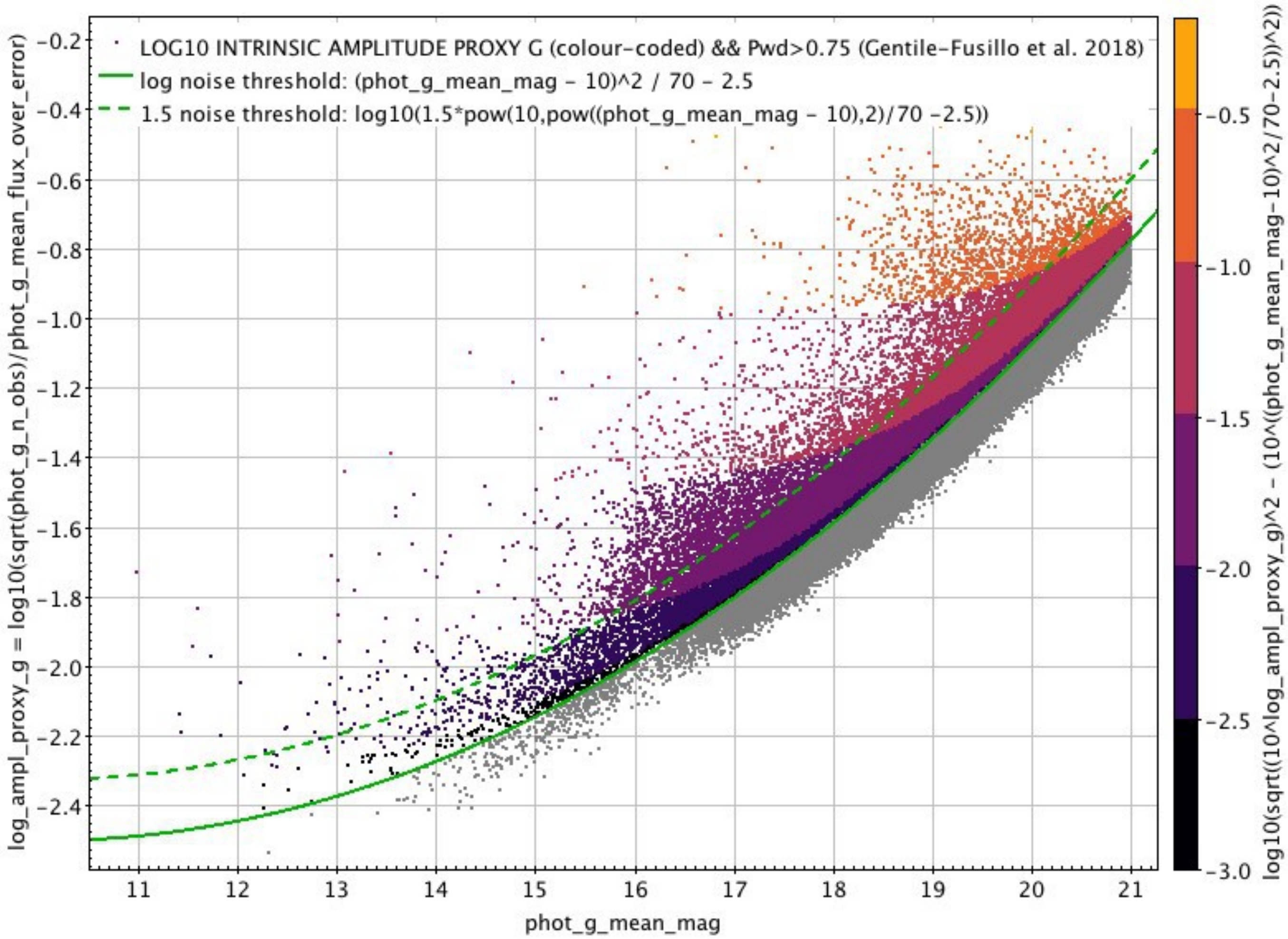}
 \caption{An amplitude proxy (defined in terms of published quantities) is shown for a selection of white dwarf candidates \citep[with probability greater than 0.75 in][]{GentileFusillo2019} as a function of the mean G-band magnitude. Intrinsic variation amplitudes were computed with respect to the assumed noise level (solid line, which represents an indicative average for constant objects) and they are colour-coded as shown in the legend. Most of the white dwarfs above the dashed green line are expected to be variable.}
 \label{eyer:fig1}
\end{figure}

\begin{figure}[ht!]
 \centering
 \includegraphics[width=0.799\textwidth,clip]{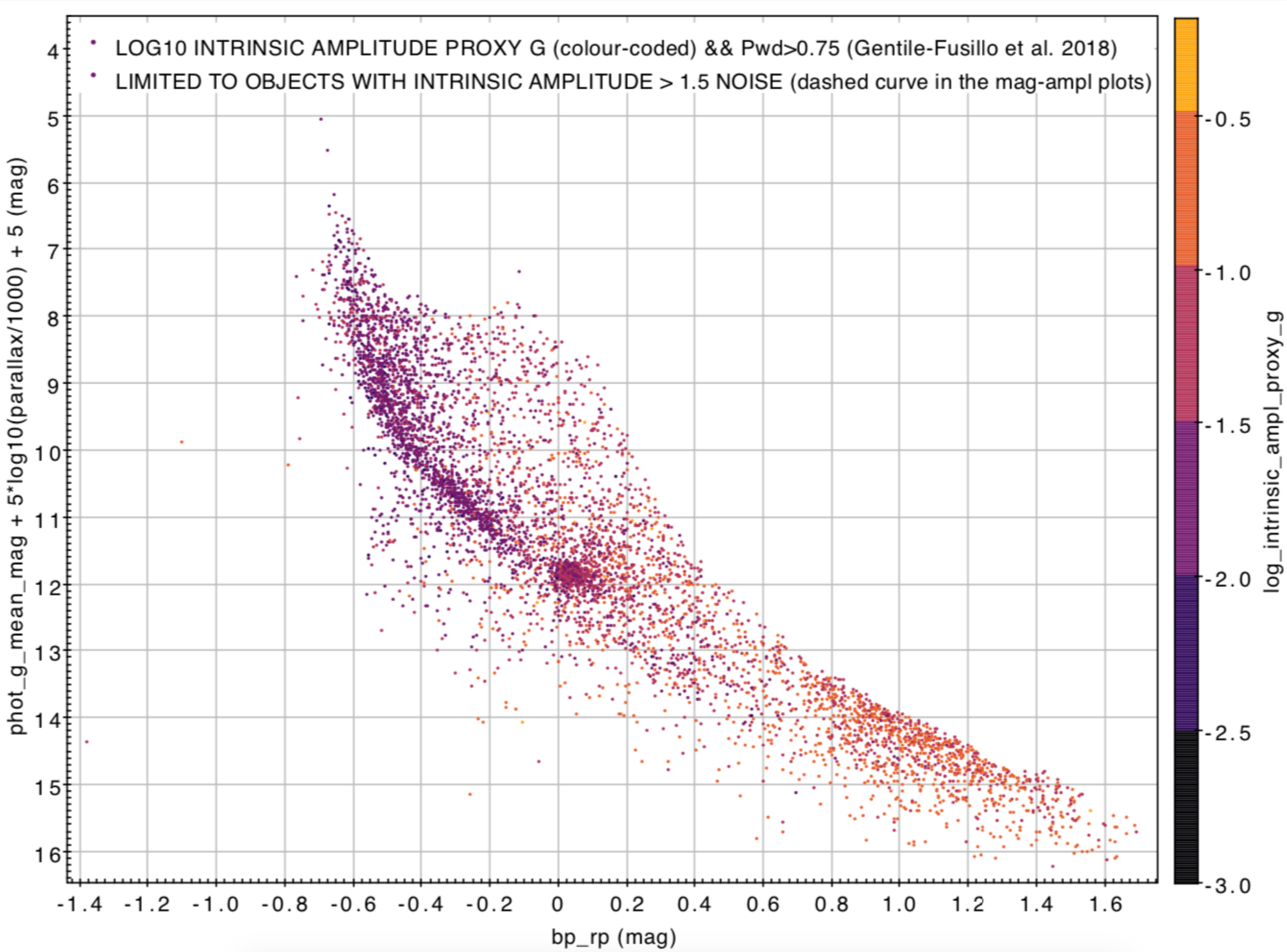}
 \caption{The observational Hertzsprung-Russell diagram of the variable white dwarf candidates. We note the presence of several regions of over-densities: those that correspond to the known ZZA, ZZB, and ZZO types (from G$_{\rm BP}-$G$_{\rm RP}\sim 0.1$~mag to bluer objects) and one at the red end that is associated with high astrometric excess noise (and thus likely spurious).}
 \label{eyer:fig2}
\end{figure}

\begin{figure}[ht!]
 \centering
 \includegraphics[width=0.799\textwidth,clip]{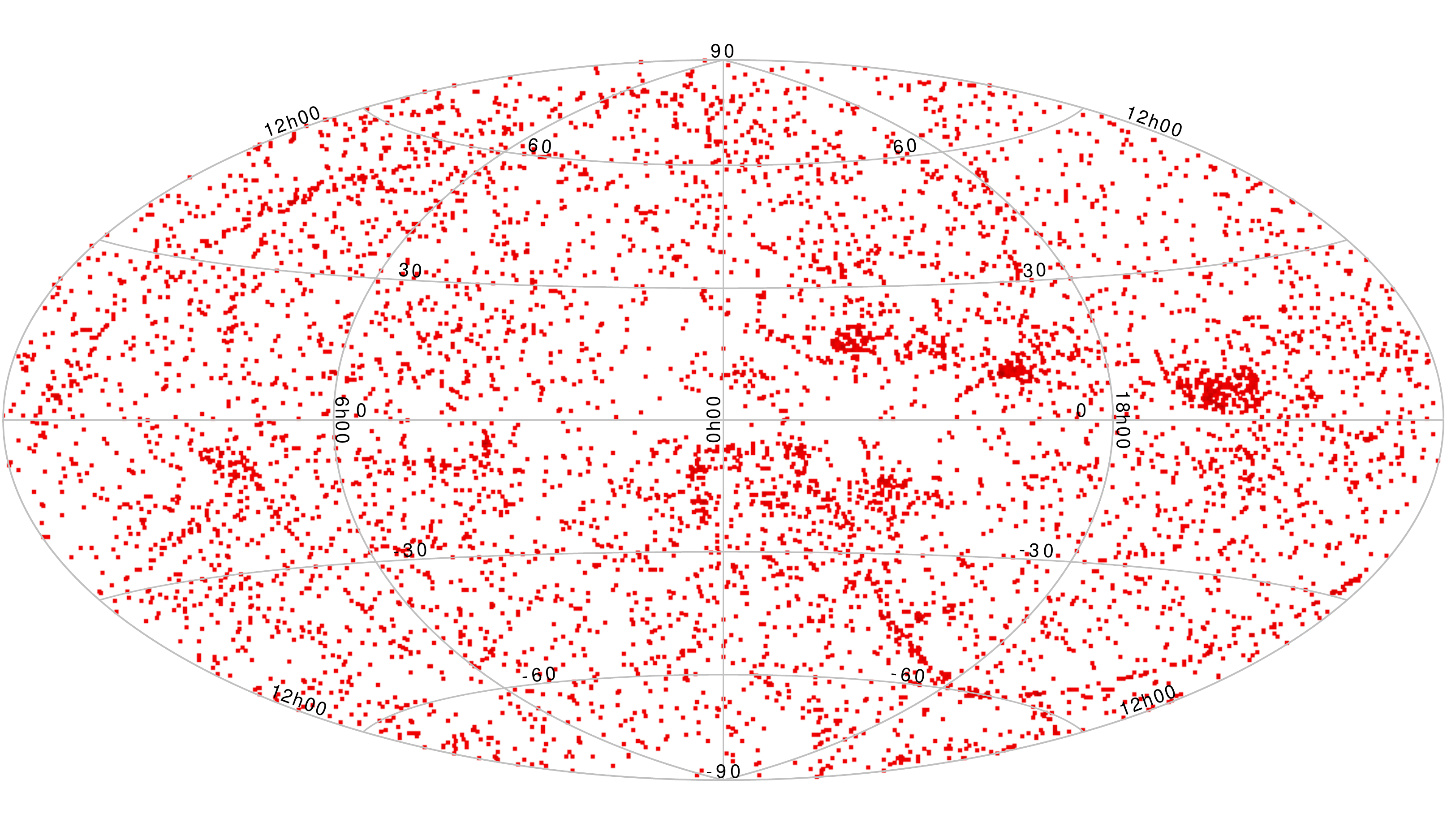}
 \caption{The sky map of variable white dwarf candidates (in Galactic coordinates). We notice some filamentary features related to the scanning law (due to outliers within given time intervals). There are also clumps of candidates in regions of low interstellar extinction, most of which correspond to the (likely spurious) over-density at the red-end in Fig.~\ref{eyer:fig2}.}.
 \label{eyer:fig3}
\end{figure}

We detected a whole new set of variable white dwarf candidates from a Gaia DR2 based selection of white dwarfs by \citet{GentileFusillo2019}. To this end, we used the published Gaia DR2 data \citep{DR2Gaia2018}, which contain averaged quantities for the photometry of all sources, with the exception of the released variable objects which include also time series, see \cite{Holl2018}. We used the uncertainty on the mean and the number of per-CCD measurements to estimate the standard deviation of the (unpublished) photometric time series.
This proxy of amplitude was then used to detect variable white dwarfs as a function of magnitude \citep[e.g.,][]{Eyer1998}, as described in Fig.~\ref{eyer:fig1}. 
A colour-absolute magnitude diagram of the variable white dwarf candidates is presented in Fig.~\ref{eyer:fig2} and the known locations of pulsating white dwarfs are clearly visible. We see the very distinctive clump of ZZ Ceti stars at BP$-$RP $= 0.05$ and M$_G = 12.2$, an elongated structure of V777 Her stars at BP$-$RP $= -0.25$ and M$_G = 11.2$ and an even more elongated regions of GW Vir stars at BP$-$RP $= -0.5$ and M$_G = 9$. The low general background  density of points, are very probably false detections. A sky map of these variable candidates is shown in Fig.~\ref{eyer:fig3} and it highlights problematic regions with likely contaminants (from the over-densities correlated in the sky). Such spurious features are well known to the consortium. Such problematic data
has been also apparent in the quasars map of \cite{BailerJones2019}, and is probably coming from their indicator of variability.
The list of sources is made available along with 
this publication and also in \url{https://www.unige.ch/sciences/astro/variability/en/data/}. This list is containing 5837 candidates among them many new true pulsating white dwarfs.

\section{The Galaxy in 3D with RR~Lyrae stars}
\begin{figure}[ht!]
 \centering
 \includegraphics[width=0.799\textwidth,clip]{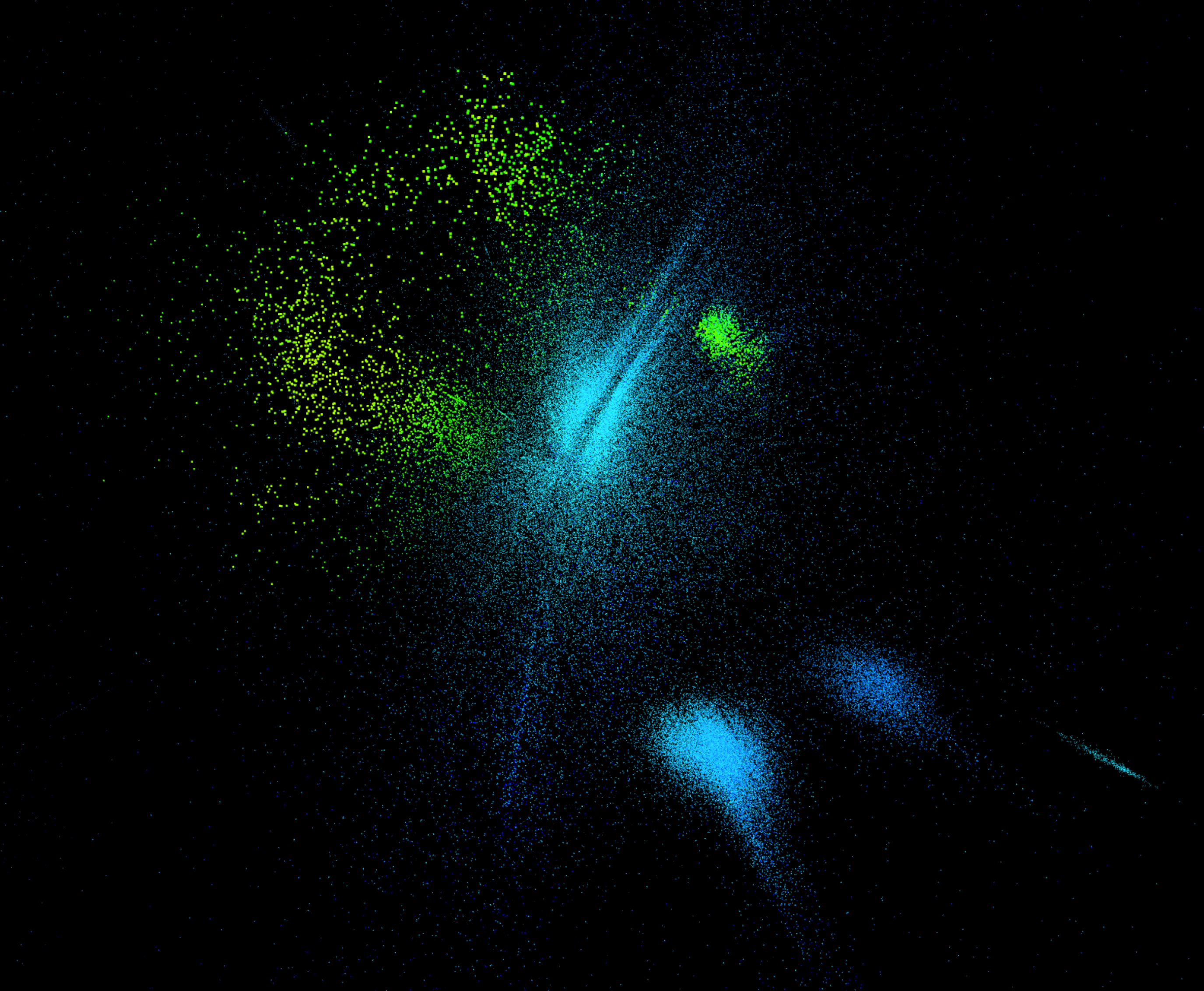}
 \caption{Distribution of 141,000 RR~Lyrae stars from \citet{Clementini2019} in Gaia DR2. The green points were selected from a specific region in the space of parallax and proper motion, and trace the Sagittarius galaxy and stream. 
}
 \label{eyer:fig4}
\end{figure}

We can obtain a view of the structures in and around the Milky Way thanks to the RR~Lyrae stars \citep[from DR2,][]{Clementini2019}. We used the simplistic assumption that the RR~Lyrae stars from this catalogue have an absolute magnitude of M$_G$ = 0.6 and derive their distance through the distance modulus, without taking into account the dependencies on extinction and metallicity. 
The 3-dimensional view in Fig.~\ref{eyer:fig4} shows the Large and Small Magellanic Clouds, the Sagittarius dwarf galaxy and the associated Sagittarius stream. We can also see small `streaks' corresponding to globular clusters. The precision of the distances is reduced by the assumptions mentioned above and causes visible radial tails, though the location of the sources in the sky is very precise. This figure can be seen as an online animation using (only) Chrome browser at the following address:
\url{https://obswww.unige.ch/~eyer/lroro/presentation/}. This animation can be also rolled back and stop at any time and can be interactively explored.

\section{Future and other data releases}
The Gaia first data release occurred in September 2016 and the second data release in April 2018. The work during these data releases was extremely intense and it became clear that an interval of 2 years between successive releases was too short. Currently, the next foreseen release is the third data release (DR3) based on 33/34 months of data. This data release will be split into:
\begin{itemize}
    \item an early data release (EDR3) in 2020, containing improved astrometry (positions, parallaxes, proper motions) and mean photometry (integrated G, G$_{\rm BP}$, G$_{\rm RP}$),
    \item the full DR3 release in 2021, including 5-10 million variable stars, with a classification, and the associated photometric time series. This release will also include: object classification and astrophysical parameters, together with BP/RP spectra and/or RVS spectra on which they are based; mean radial velocities for stars with available atmospheric-parameter estimates; solar-system results with preliminary orbital solutions and individual epoch observations; non-single stars; quasars and results for extended objects. In addition, we proposed to release a pencil beam with epoch photometry for all sources (variable and non-variable), centered on the Andromeda Galaxy.
\end{itemize}
The latest updates and details on the Data Releases are available on the ESA webpage\footnote{ \url{https://www.cosmos.esa.int/web/gaia/release}}.

The 4th data release is not fixed but may happen in 2024, the schedule is being currently discussed within the consortium. This release should contain the first 5 years of data (the nominal mission) and possibly a part of the data from the extended mission.

A final data release containing the nominal mission and the data from the extension (at most 5 years) will be released (though it is too early to specify a date).

\section{Conclusion}
By essence/design Gaia is a time domain space machine.
Gaia time series analysis produces the parallaxes, proper motions, and various photometrically and spectroscopically variable signals. Their nature can be periodic, quasi-periodic, transient or stochastic. These signals with Gaia animates directly many diagrams as shown in this contribution, however they are just the shiny tip of the iceberg of Gaia's contribution to astronomy --- Gaia has and will have a staggering impact in nearly all domains of astrophysics, also thanks to its content of variable celestial objects.


\begin{acknowledgements}
Acknowledgement:
We would like to thank late Prof.~Gilles Fontaine for the interactions and encouragements on the variable white dwarf candidates form Gaia. This work has made use of data from the European Space Agency (ESA) mission
{\it Gaia} (\url{https://www.cosmos.esa.int/gaia}), processed by the {\it Gaia}
Data Processing and Analysis Consortium (DPAC,
\url{https://www.cosmos.esa.int/web/gaia/dpac/consortium}). Funding for the DPAC
has been provided by national institutions, in particular the institutions
participating in the {\it Gaia} Multilateral Agreement.
\end{acknowledgements}

\bibliographystyle{aa} 
\bibliography{Eyer_2k01.bib} 

\end{document}